\documentclass[aps,prb,twocolumn,groupedaddress,floatfix,showpacs]{revtex4-1}
\usepackage[T1]{fontenc}
\usepackage{graphicx}
\usepackage{amsmath}
\usepackage{subcaption}

\usepackage{color}
\newcommand{\parl}{\parallel}
\begin{document}

%\preprint{}

\title{Optical properties of Rydberg excitons in Cu$_2$O based superlattices}


\author{David Ziemkiewicz}
\email{david.ziemkiewicz@utp.edu.pl}

%\thanks{}
\author{Gerard Czajkowski}
%\author{Karol Karpi\'{n}ski}
\author{Sylwia Zieli\'{n}ska-Raczy\'{n}ska}

 \affiliation{Institute of
Mathematics and Physics, UTP University of Science and Technology,
\\ Al. Prof. S. Kaliskiego 7, 85-789 Bydgoszcz, Poland}


\date{\today}

\begin{abstract}  Combining   the  microscopic
calculation of superlattice minibands   and the macroscopic  real
density matrix approach one can obtain electric
susceptibilities of the superlattice system irradiated by an electromagnetical wave. It is shown how to compute the dispersion relation, excitonic resonances positions and susceptibility
of Cu$_2$O/MgO based superlattice (SL), when {Rydberg
Exciton-Polaritons} appear, including the effect of the coherence
between the electron-hole pair and the electromagnetic  field and
the polaritonic effect. Using the Kronig-Penney model  for computing miniband SL parameters the analytical expressions for
optical functions are obtained and the numerical calculations for Cu$_2$O/MgO SL are performed.
\end{abstract}

%\keywords{}

\maketitle

\section{Introduction}
Excitons, Coulomb-bound pairs of a one conduction band electron and a one valence band hole, form an electrical neutral quasiparticle, transferring the energy without transporting the net electric charge \cite{Klingshirn}. These quasiparticles are complex many-body states embedded in the background of crystal lattice, which interact via scattering, phase-space and screening.
Therefore the problem of a manipulation of exciton states through application of artificial periodic potentials systems has attracted lot of attention; some implementations include colloidal semiconductor nanocrystals \cite{Harankahage}, microrod arrays \cite{Zhang18} and micropillars \cite{Ghosh}. As pointed out in \cite{Zang2017}, so-called structured excitons can be used as a means of transporting information and energy in quantum information processing. One of the ways to control these excitons is via superlattice \cite{yang}. Such  a system causes a large shift of exciton energy states and thus influences optical and electronic properties. In principle, SL containing Rydberg exciton is a solid-state analogue of Rydberg atom trapped in an optical lattice, which are a promising tool in quantum computing \cite{Isenhower2010,Omran2019,Li2020}. Superlattice can also be used as a medium for exciton-exciton interaction experiments \cite{Zang2017}. Importantly, superlattice has an advantage of relative simplicity and ease of fabrication. Moreover, in the case of Cu$_2$O/MgO system proposed here, when low principal number excitons are used, room temperature operation is feasible \cite{Mazanik2022}.

A superlattice  is a periodic structure of layers made of two (or more) semiconductor or insulator materials with different band gaps, each quantum well sets up new selection rules that affect the conditions for charges to flow through the structure. The two different semiconductor materials are deposited alternately on each other to form a periodic structure in the growth direction. Typically the width of layers is order of magnitude larger than the lattice constant, and is limited by the growth of the structure. Due to the small width of individual layers, on the scale of illuminating light wavelength they merge together to form a homogeneous system, which behaves like a bulk crystal. Important requirements of producing a superlattice  are a small lattice mismatch and different band gaps energies between two material components of the structure.

Recently  Yang \textit{et al.}\cite{yang} produced SL based on cuprite, where the wells consist of a narrow-bandgap semiconductor Cu$_2$O and the barriers are made of a wide-bandgap insulator MgO. The lattice constants of both these substances are quite similar with a small mismatch between the constituent layers (with the difference 1.35 $\%$), while Cu$_2$O is a narrow-band gap  semiconductor with E$_g\sim$ 2.2 eV and MgO is a wide-band gap compound with E$_g\sim$ 8 eV  \cite{yang}, which satisfy the basic requirements for a good SL structure. 

In this paper, we consider a structure of similar dimensions, e.g. total thickness on the order of 100 nm and individual layer thickness on the order of few nm. We intend to describe
optical properties of this SL:  the optically active layers of cuprous oxide Cu$_2$O and buffer layer of magnesium oxide MgO. 
 In our  paper we will discuss the behavior of Rydberg excitons located in the system of quantum wells, which create a system of periodic potentials. Since the
first observation of Rydberg excitons (REs) in Cu$_2$O in 2014
\cite{Kazimierczuk}, they become a subject of intensive studies. These  highly excited states in  Cu$_2$O,  were observed up to a large principal quantum number $n=30$ \cite{Versteegh}. Due to unusual properties of REs, such as huge sizes scaling as $n^2$, long life times reaching nano seconds, strong exciton-exciton interactions controlled by so-called Rydberg blockade, REs could have many promising applications as single-photon emitters, single photon transistors and as active medium of masers. \cite{DZ_OL}
Initial studies on Rydberg excitons were focused at the optical properties of
REs in high quality nature crystals (bulk crystals), see \cite{Stolz}$^,$\cite{Morin} for recent references. Also some groups
concentrated on  fabrication techniques of Cu$_2$O nanostructures \cite{Steinhauer, Takahata}. 
Recently, the main  interest of research has shifted from REs
 in bulk crystals to excitons in low-dimensional systems
\cite{Konzelmann}$^-$\cite{Ziemkiewicz_2023}. The first experimental verification  of an oscillator strength change caused by  the quantum confinement of REs in low dimensional quantum system \cite{Orfanakis} was an important step forward to exploit them in quantum applications.  
 Therefore it seems natural to examine the optical properties of REs in the specific type of a nanosystem, which consists of quantum wells forming lattice of periodic potentials confining REs.
The unique property of periodic potential systems is a possibility of changing effective masses of particles inside such structures. Regarding an exciton in SL, an electron and a hole effective masses are modified, which results in adjustment of an optical susceptibility. For a given SL structure geometry one is able to predict the shift of excitonic resonances positions comparing to the bulk case.
 The last but not the least argument for choosing this subject is the fact that 
the optical lattices with neutral atoms have been successfully applied in quantum information devices. In  analogy we imply that due to the inherent, repeating pattern of  and long-coherence times of Rydberg excitons, their huge polarizability and dipole moments, which allow them strongly interact with each other over a long distance, arranged in such systems, they might be also viable candidates for quantum computing.

  Band-edge optical
properties of superlattices can be discussed by modelling the
superlattice as an effective anisotropic medium in which the
quasi-free carriers propagate and interact. In the low barriers
limit the electron and hole motion in the confinement direction is
determined by the superlattice potential and is replaced by an
effective-mass motion, with the appropriate effective masses
obtained from the miniband dispersion relations
\cite{Bastard,Pereira}.

Since excitons in the majority  of semiconductors are of Wannier type, the
transition dipole  has a spatial extension, characterizing
the interaction of radiation with electrons and holes located at
different sites. This results in a coherence between the electron-hole
pair and the radiation field. In analogy to bulk semiconductor
excitons, SL excitons induced by an electromagnetic
wave propagating through the SL will give rise to
``$SL-polaritons$''. 

As in the bulk crystals, polaritons are mixed modes of the electromagnetic field
and discrete excitations of the SL $E_n({\bf k}_{\rm ex})$ (excitons). Below the gap one can imagine a polariton as a photon surrounded by a cloud of virtual electron-hole pairs (excitons).

All the above mentioned components (Wannier excitons,
effective mass approximation, exciton-polaritons with coherence)  
justify the use of the Real Density Matrix Approach (RDMA)  to
describe optical properties of superlattices. The method has been
already used to describe  excitons and polaritons in
III-V\cite{CBT_1996} and II-VI SL\cite{Schillak} and was 
successful in description of REs optical properties of Cu$_2$O bulk
crystals\cite{Zielinska.PRB},  and nanostructures (quantum wells, dots and wires)\cite{Ziemkiewicz_QD}.

Below we present in details a procedure of calculation, which starts with the Kronig-Penney model to obtain SL miniband parameters i.e.,
 anisotropic effective masses and band gaps. To derive the dispersion relation and resonance positions in SL, RDMA with these parameters is used. This method has general character, allows to get analytical formula for a system susceptibility. It takes into account both the
Coulomb interaction between an electron and a hole and coherence between an electron-hole pair and a radiation field.
The particular calculations will be done for Cu$_2$O/MgO SL, for which the SL
dielectric tensor and the optical functions in the analytical form will be calculated. 

The paper is
organized as follows. In Sec. II we present the basic equations of
the Kronig-Penney model adapted to the cases of superlattices.  
Sec. III shows the scheme for calculating SL optical functions
in the case when the total thickness of the SL is much greater
than the excitonic Bohr radius. In Sec. IV results
obtained for Cu$_2$O/MgO superlattice are discussed and  conclusions are presented in Sec V.
\section{Kronig-Penney model for superlattices}\label{sec2}
In this section we recall the basic equations which describe the
electronic states (conduction and valence bands) of a
superlattice. Considering the Kronig-Penney model we assume the confinement
potential in the $z$-direction (structure growth direction), which for conduction electrons
corresponds to $V(z)=0$ if $z$ corresponds to area inside well, (well thickness
$L_W$, effective mass $m_W$), and $V(z)=V_0$ if
$z$ corresponds to the barrier area (thickness $L_B$, effective mass $m_B$),
where $V_B$ is the conduction band offset. The equation for
the values of the Bloch vector $K$, and thus the miniband
dispersion \cite{Bastard}$^-$\cite{Davies}, takes the
Kronig-Penney form  
\begin{eqnarray}\label{KPbound}
&&\cos KL=\cos k_1L_W\cosh \kappa_2L_B\nonumber\\
&&-\frac{k_1^2-\kappa_2^2}{2 k_1\kappa_2}\sin k_1L_W\sinh
\kappa_2L_B,
\end{eqnarray}
\noindent where $k_1$ and $\kappa_2$ are the wave vectors in the wells, and in the barrier, respectively. The subscripts $W$ and $B$ in Eqs.~(\ref{KPbound}) denote the wells or barriers, and
$L=L_W+L_B$ is the SL period. The wave vectors in the well an barrier are
\begin{eqnarray}
&&k_1=\sqrt{\frac{2m_W E}{\hbar^2}}\nonumber\\
&&\kappa_2=\sqrt{\frac{2m_B (V-E)}{\hbar^2}}.
\end{eqnarray}
The above equations can be solved for electrons and holes separately, obtaining the relation $E(K)$ where $E$ is the electron/hole energy, $m_W$ and $m_B$ are effective masses in Cu$_2$O and MgO, and $V$ is the potential barrier between MgO and Cu$_2$O. Specifically, due to the difference of band gap energies (see Table \ref{parametervalues}), we have $V_0=4.99$ eV. From this, we obtain the electron and hole confinement potentials $V_{0e,h}$
\begin{eqnarray}\label{confinement}
V_0&=&E_g(\hbox{MgO})-E_g(\hbox{Cu$_2$0}),\nonumber\\
V_0&=&V_{0e}+V_{0h},\\
V_{0e}&=&0.4\times V_0,\qquad V_{0h}=0.6\times
V_0.\nonumber\end{eqnarray}
This results in two barrier values $V_{0e}=2$  eV and $V_{0h}=3$ eV for electrons and holes, respectively. The division of $V_0$ into electron and hole potential barriers follows from the relative Fermi energy level of Cu$_2$O \cite{Balamungan2004} and MgO \cite{VanHuis2003} and is similar to the case of Cu$_2$O/ZnO heterojunction \cite{Siol2016}.
 
One of the important parameters of SL structure that differentiates it from bulk material are the effective masses of a hole and an electron in the z direction, which are determined from the relation
\begin{equation}\label{eq:masy}
\frac{1}{m_z}=\left.\frac{1}{\hbar^2}\frac{d^2E}{dK^2}\right|_{K=0}.
\end{equation}
which, again, can be obtained separately for electrons and holes, from the respective $E(k)$ relations.

\section{Optical properties}\label{sec3}
\begin{figure}[ht!]
\centering
\includegraphics[width=.6\linewidth]{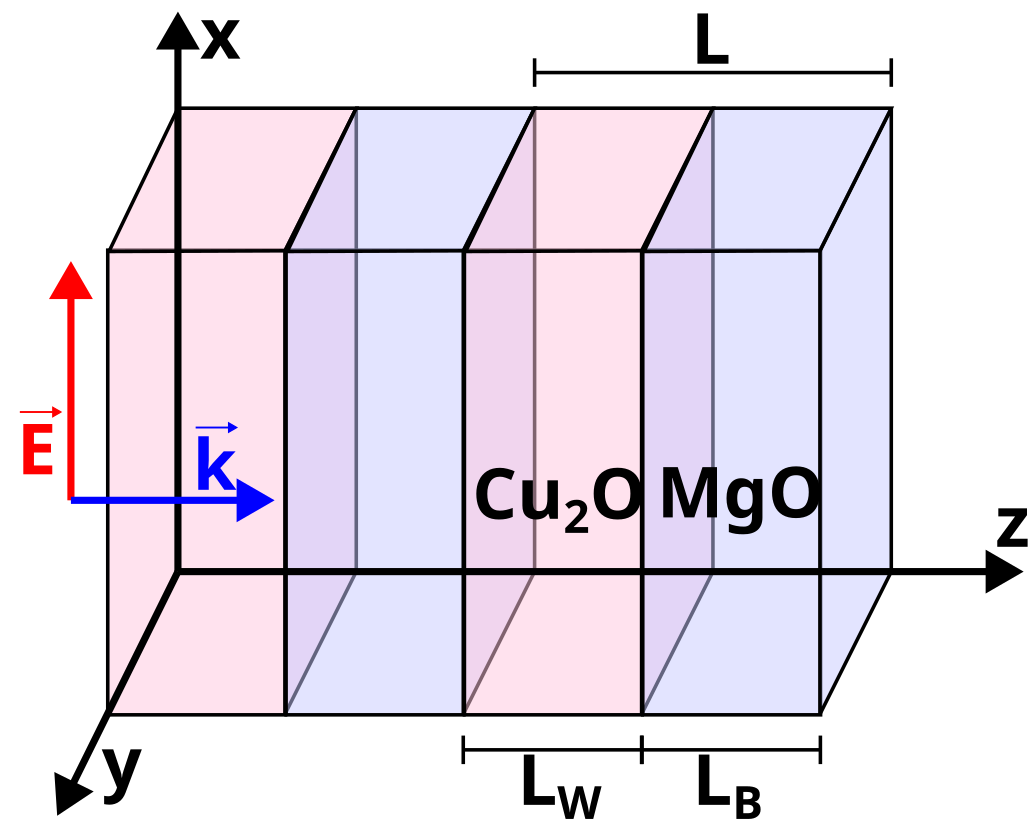}
\caption{Schematic representation of the system.}\label{rys:schemat}
\end{figure}

We consider a superlattice consisting of multiple layers of wells and barriers, characterized by a thickness $L_W$ and $L_B$ and a total thickness of a single well-barrier pair $L$. The system is presented in Fig. \ref{rys:schemat}. It is irradiated by a normally incident electromagnetic wave, linearly polarized in the $x$-direction   
\begin{equation}\label{incidentfield}
E_i(z,t)=E_{i0}\exp(ik_0z-i\omega t), \quad k_0=\frac{\omega}{c}.
\end{equation}
We assume that $L<4$ nm. The total number of layers is on the order of 10-100, and the exact number is not relevant to the calculations.

The linear optical response of the system (here we consider the lowest electron and hole miniband) to the electromagnetic wave
originates from a given pair of minibands, and is described by two
equations: the so-called constitutive equation (material equation) and the Maxwell's propagation propagation equation.
 The constitutive equation has the form
\begin{eqnarray}\label{constitutiveHL}
&& -i\hbar \partial_tY -i\Gamma Y+ H_{eh}Y= {\bf M}({\bf r}){\bf
E}({\bf R}),
\end{eqnarray}
\noindent where $Y(\textbf{R},\textbf{r},t)$ is the excitonic
transition coherent amplitude, $\Gamma$ is a dissipation coefficient,
${\bf M}$ is the transition dipole density, $\textbf{R}$ is the
excitonic center-of-mass coordinate, and $\textbf{r}$ the relative
electron hole-coordinate. The operator $H_{eh}$ is the  effective
mass  Hamiltonian of the superlattice

\begin{equation}\label{SLehHamiltonian}
H_{eh}=E_{g}+\frac{P_Z^2}{2M_{z}}+
 \frac{{\bf P}_{\parl}^2}{2M_{\parl}}
+\frac{p_z^2}{2\mu_{z}}+ \frac{{\bf p}_{\parl}^2}{2\mu_{\parl
}}+V_{eh},
\end{equation}
\noindent with $V_{eh}$ being the electron-hole Coulomb interaction. We have
separated the center-of-mass coordinate ${\bf R}_{\parl}$  and the
related momentum ${\bf P}_{\parl}$ from the relative coordinate
$\rho$ on the plane $x-y$ and the related momentum ${\bf
p}_{\parl}$. In the above formulas the reduced mass in the
$z$-direction is given by
\begin{equation}\label{mureduced}
\frac{1}{\mu_{z}}=\frac{1}{m_{ez}}+\frac{1}{m_{hz}},
\end{equation}
\noindent where the electron- and the hole effective masses in the
$z$-direction follow from the miniband dispersion relations (\ref{KPbound}), one for electrons and one for holes, respectively. The system is not confined in $xy$ directions and so the in-plane effective masses $m_\parl$ in the well material are assumed to be the same as in bulk medium.  $M_z$ and $M_\parl$ are the total
excitonic masses in the growth direction and parallel to the
layers, respectively. We use the same form for the transition
dipole density, as for bulk semiconductor \cite{Zielinska.PRB}
\begin{eqnarray}\label{gestoscwzbronione}
{\bf M}({\bf r})&=&
\textbf{e}_r\,M_{10}\frac{r+r_0}{2r^2r_0^2}e^{-r/r_0}=\textbf{e}_r
M(r) \nonumber\\
&=&\textbf{i}M_{10}\frac{r+r_0}{4{\rm
i}r^2r_0^2}\sqrt{\frac{8\pi}{3}}\left(Y_{1,-1}-Y_{1,1}\right)e^{-r/r_0}\nonumber\\
&&+\textbf{j}M_{10}\frac{r+r_0}{4r^2r_0^2}\sqrt{\frac{8\pi}{3}}\left(Y_{1,-1}+Y_{1,1}\right)\,e^{-r/r_0}\nonumber\\
&&+\textbf{k}M_{10}\frac{r+r_0}{2r^2r_0^2}
\sqrt{\frac{4\pi}{3}}Y_{10}e^{-r/r_0},
\end{eqnarray}
where $r_0$ is the so-called coherence radius
\begin{equation}\label{r0}
r_{0}^{-1}=\sqrt{\frac{2\mu }{\hbar^2}{E_g}}.
\end{equation}
 The above expression gives the
coherence radius in terms of
 effective band parameters $E_g$ (the bulk gap energy), and $\mu$
 (the electron-hole reduced effective mass, the bulk effective masses of the electron and the hole are assumed to be isotropic). $M_{10}$ is the integrated dipole strength. 
In order to present detailed derivation of  susceptibility with Rydberg excitons adapted for a case of a superlattice we recall the procedure similar to that presented in \cite{Zielinska.PRB}.  
   The  steps of the calculation scheme are the following:
\begin{enumerate}
\item The excitonic amplitude $Y$ is determined from Eq.
(\ref{constitutiveHL}) with the Hamiltonian
(\ref{SLehHamiltonian}). 
\item  The coherent amplitude $Y$ enables to calculate the SL polarization which is given by the forumla \cite{Zielinska.PRB} 
\begin{equation}\label{polarization}
\textbf{P}(\textbf{R})=2\int d^3r {\bf M}({\bf \textbf{r}})
Y(\textbf{R},r).\end{equation} \item
The polarization $\textbf{P} $ is then inserted into the Maxwell propagation equation
\begin{equation}\label{Maxwell}
c^2\hbox{\boldmath$\nabla$}_R^2
\textbf{E}-\underline{\underline{\epsilon}}_b\ddot{\textbf{E}}(\textbf{R})=\frac{1}{\epsilon_0}\ddot{\textbf{P}}(\textbf{R}),
\end{equation}
with the use  of the bulk dielectric tensor
$\underline{\underline{\epsilon}}_b$ and the vacuum dielectric
constant $\epsilon_0$.\
\end{enumerate}
In analogy to bulk crystals, the description of SL exciton-polaritons is based on the separation of the relative electron-hole motion with well- defined quantum levels and the center-of-mass motion which interacts with the radiation field and produces the mixed modes (polaritons). We assume that the center-of-mass motion is described by the term $\exp({\rm i}\textbf{k}\,{\textbf{R}})$ with the wave vector $\textbf{k}$. Additionally, we use the EM wave that has a harmonic time dependence $\propto \exp(-{\rm i}\omega t)$. These simplifications allow us to calculate the dielectric susceptibility. Because in Cu$_2$0 the conduction band and the valence band are of the same parity the dipole moment between them vanishes; the $n>1$ lines correspond to excitons with the relative angular momentum $l=1$ therefore the absorption process is dipole-allowed.
  The Eq.
 (\ref{constitutiveHL}) will be solved
 by expanding the coherent amplitude $Y$ in
terms of eigenfunctions of the Hamiltonian $H_{eh}$,
\begin{equation}\label{expansion}
Y=\sum\limits_{n\ell m}c_{n\ell m}R_{n\ell m}(r)Y_{\ell
m}(\theta,\phi),
\end{equation}
where $n,l,m$ are main, relative momentum and magnetic quantum numbers respectively,  $Y_{\ell m}$ are spherical harmonics, which are real valued functions of the spherical coordinates $\theta,\phi$. Specifically, we use the definition
\begin{equation}\label{spherical}
Y_{\ell,m}(\theta,\phi)=\sqrt{\frac{2\ell+1}{4\pi}\;\frac{(\ell-m)!}{(\ell+m)!}}\;P_{\ell}^m(\cos\theta)e^{im\phi},
\end{equation}
where $P_{\ell,m}$ are the associated Legendre polynomials,
\begin{equation}\label{Legendre}
P_\ell^m(x)=\frac{(-1)^m}{2^\ell\,\ell!}\left(1-x^2\right)^{m/2}\frac{d^{\ell+m}}{d\,x^{\ell+m}}\left(1x^2\right)^\ell.
\end{equation}

The radial functions $R_{n\ell m}$ are given in the form
\begin{eqnarray}\label{radialfinala}
&&R_{n\ell m}(r)=\left(\frac{2\eta_{\ell
m}}{na^*}\right)^{3/2}\frac{1}{(2\ell
+1)!}\sqrt{\frac{(n+\ell)!}{2n(n-\ell-1)!}}\\
&&\times\left(\frac{2\eta_{\ell m}r}{na^*}\right)^\ell
e^{-\eta_{\ell m}r/na^*}M\left(-n+\ell+1,2\ell+2,\frac{2\eta_{\ell
m}r}{na^*}\right).\nonumber
\end{eqnarray}
The coefficient $\eta_{lm}$ depends on an effective masses ratio $\alpha$, which for SL is different from the bulk and therefore is crucial for eigenvalues  $E_{n\ell m}$
\begin{equation}
\eta_{\ell m}=\int {
d}\Omega\frac{\vert Y_{\ell m}\vert^2}{\sqrt{\sin^2\theta
+\alpha\cos^2\theta}},
\end{equation}
\begin{eqnarray}\label{energies}
E_{n\ell m}&=&-\frac{\eta^2_{\ell m}(\alpha)R^*}{n^2},\quad
n=1,2,\ldots,\nonumber\\
&\quad& \ell=0,1,2,...n-1,\quad m=0,1,2,...\ell,\nonumber\\
&&\alpha={\frac{\mu_\parl}{\mu_z}}.
\end{eqnarray}
 $a^*$ is
 the exciton Bohr radius, $M(a,b,z)$ is the Kummer function
 (confluent hypergeometric function) in the notation of Ref.
 \cite{Abramowitz}. The anisotropy parameter $\alpha$, first introduced by
Kohn and Luttinger \cite{Kohn}, corresponds to the dimensionality of the system \cite{Mathieu,Czajk}; specifically in the so-called Fractional Dimensionality Approach \cite{He90}, the system dimension $d=2+\sqrt{\alpha}$, so that it is two-dimensional in the limit of $\alpha \rightarrow 0$ and 3-dimensional for $\alpha=1$.

\noindent $R^*$ is the effective excitonic Rydberg energy defined as
\begin{equation}\label{excitonicRydberg}
R^*=\frac{\mu_{\parl}e^4}{2(4\pi\epsilon_0\sqrt{\epsilon_{\parl}\epsilon_z})^2\hbar^2}.
\end{equation}
With the modified by periodic potential of SL effective masses, one can calculate the anisotropy factor $\alpha$  and the
corresponding eigenvalues $E_{n\ell m}$ from eq.(\ref{energies}). In the
considered case of $P$ excitons we use the quantities
\begin{eqnarray}\label{etaellm}
\eta_{00}(\alpha)&=&\frac{\arcsin\sqrt{1-\alpha}}{\sqrt{1-\alpha}},\nonumber\\
\eta_{10}(\alpha)&=&\frac{3}{2(1-\alpha)}\left(\eta_{00}-\sqrt{\alpha}\right),\\
\eta_{11}(\alpha) &=&\frac{3}{2}\left[\eta_{00}(\alpha)
-\frac{1}{3}\eta_{10}(\alpha)\right].\nonumber
\end{eqnarray}
The energies for electron and hole $E_{0e,h}=E_{e,h}(k=0)$ determine the SL energy gap
\begin{equation}\label{SLgap}
E_g(\hbox{SL})=E_g +E_{0e}+E_{0h},
\end{equation}
which is shifted and, together with the eigenenergies $E_{n\ell m}$, the positions
of SL excitonic resonances, given by the transverse energies $E_{Tn\ell m}$
\begin{equation}\label{ET}
E_{Tn\ell m}=E_g(\hbox{SL})+E_{n\ell m},
\end{equation}
which are also moved due to the modification of effective masses by the influence of SL periodic potentials pattern.

Since the superlattice consists of multiple quantum wells, it is justified to assume that  symmetry properties of excitons in SL are similar to these in quantum wells. The considered system geometry (see Fig. 1) and the
electric field polarization (\ref{incidentfield}) allows to use the
\textbf{i}-component of the dipole density
(\ref{gestoscwzbronione}),
\begin{equation}\label{Mx}
M_x(r)=M_{10}\frac{r+r_0}{4{\rm
i}r^2r_0^2}\sqrt{\frac{8\pi}{3}}\left(Y_{1,-1}-Y_{1,1}\right)e^{-r/r_0}.\end{equation}
With the help of Eqs. (\ref{radialfinala}) and (\ref{Mx}) 
the expansion coefficients $c_{n\ell m}$ are calculated. Then the coherent
amplitude $Y$ is used in Eq. (\ref{polarization}), which in turn
is inserted into the Maxwell equation (\ref{Maxwell}), from which
one obtains the dispersion relation for SL-polaritons
\begin{eqnarray}\label{dispersion}
&&\frac{c^2k^2}{\omega^2}-\epsilon_b\\
&&=\epsilon_b\sum\limits_{n=2}^N\frac{\Delta_{LT}^{(P)}f_{n11}}
{E_{Tn11}-\hbar\omega+(\hbar^2k^2/2M_z)-i{\mit\Gamma_n}},\nonumber
\end{eqnarray}
with
\begin{equation}\label{eq:fn11}
f_{n11}=\frac{32}{3}\frac{(n^2-1)\eta_{11}^5}{n^5}.
\end{equation}
 $\Delta_{LT}^{(P)}$ is the longitudinal-transversal splitting
energy, and $E_{Tn11}$ are the energies of excitonic resonances
(see Eq. (\ref{ET})). The relation \cite{Magnetoexcitons} 
\begin{equation}
\left|M_{10}\right|^2=\frac{4\epsilon_0\epsilon_b
a^{*3}\Delta_{LT}^{(P)}}{\pi (r_0/a^*)^2},\nonumber
\end{equation}
has been used. In the considered, narrow frequency range, one can use $\epsilon_b=const=7.5$ \cite{Kazimierczuk}.

The spatial dispersion, described by (\ref{dispersion}), makes it possible to have two or more
wave modes connected with a given frequency. The term $\hbar^2k^2/2M_z$ in the denominator on
r.h.s. of Eq. (\ref{dispersion}) is responsible for the effect of
multiplicity of polariton waves. In the considered case of Cu$_2$O/MgO
SL the total exciton mass $M_z$ is much larger than in other
semiconductors (both bulk and SL) so it is justified to neglect this term.
As pointed out in \cite{Orfanakis22}, relatively small oscillator strength and resulting weak light-exciton coupling in bulk Cu$_2$O makes it difficult to achieve strong coupling regime necessary for polaritonic effects to become significant. It is demonstrated in \cite{Orfanakis22} that this problem can be solved by placing the crystal between two Bragg reflectors, forming a cavity; multiple, strong reflections on Cu$_2$O/MgO interfaces considered here might provide another way of achieving strong coupling regime. In view of the above findings, we obtain the excitonic contribution to the linear optical susceptibility in the form $\chi(\omega)$,
\begin{equation}\label{susceptibility}
\chi(\omega)=\epsilon_b\sum\limits_{n=2}^N\frac{\Delta_{LT}^{(P)}f_{n11}}{
E_{T n 1 1}-\hbar\omega-i{\mit\Gamma_n}}.
\end{equation}
In particular, we are interesten in the imaginary part of susceptibility, where absorption maxima corresponding to excitonic states can be observed; we note that apart from these maxima, both Cu$_2$O and MgO are mostly transparent, with significant abosrption occuring only on a length scale of tens of $\mu$m. 

\section{Results}
The above presented scheme allows the calculation of all optical SL functions. We have chosen the optical susceptibility since its imaginary part is proportional to the SL absorption. We have computed the susceptibility Cu$_2$O/MgO SL for a variety of Cu$_2$O QW and MgO barrier thicknesses. The values of the relevant parameters are given in Table \ref{parametervalues}. The obtained results are illustrated in Figures 2-9.

\begin{figure}[ht!]
\includegraphics[width=.7\linewidth]{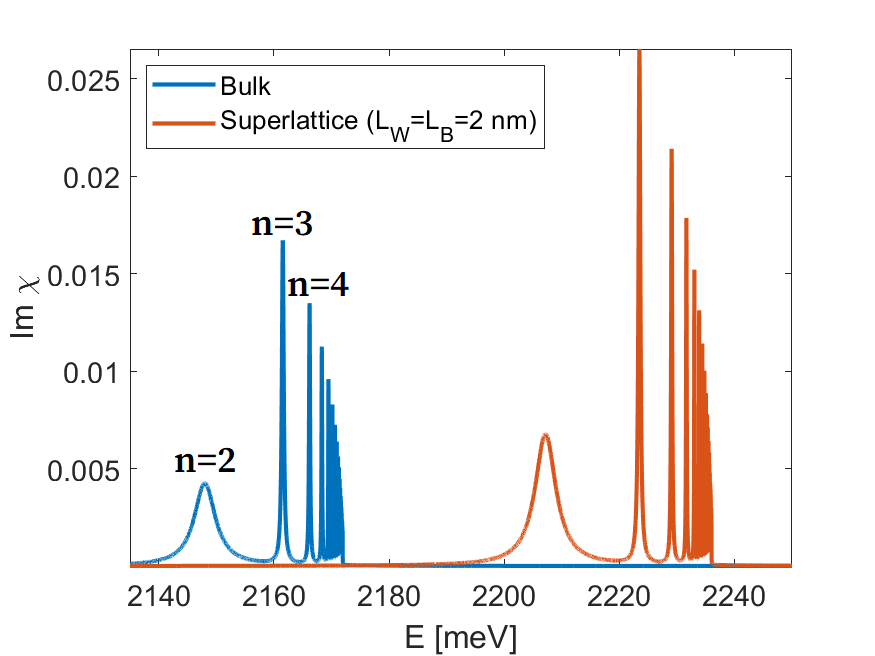}
\caption{Comparison of the imaginary part of susceptibility in bulk Cu$_2$O and in Cu$_2$O/MgO SL, calculated from Eq. (\ref{susceptibility}). First few excitonic states $n=2..4$ are marked.}\label{rys:1}
\end{figure}

The Fig. \ref{rys:1} presents the imaginary part of susceptibility of bulk Cu$_2$O and a superlattice with L=4 nm ($L_W=L_B=2$ nm). Even for such a relatively large L (on the order of 10 lattice constants)), the energy shifts $E_{e0}$, $E_{h0}$ are considerably larger than energy spacing of excitonic levels.  
The two key features visible in Fig. \ref{rys:1} is a slight increase of oscillator strength (proportional to the area under the absorption peak) is SL, as well as a slight modification of Rydberg energy, in accordance with  Eq. (\ref{excitonicRydberg}).

\begin{figure}[ht!]
a)\includegraphics[width=.8\linewidth]{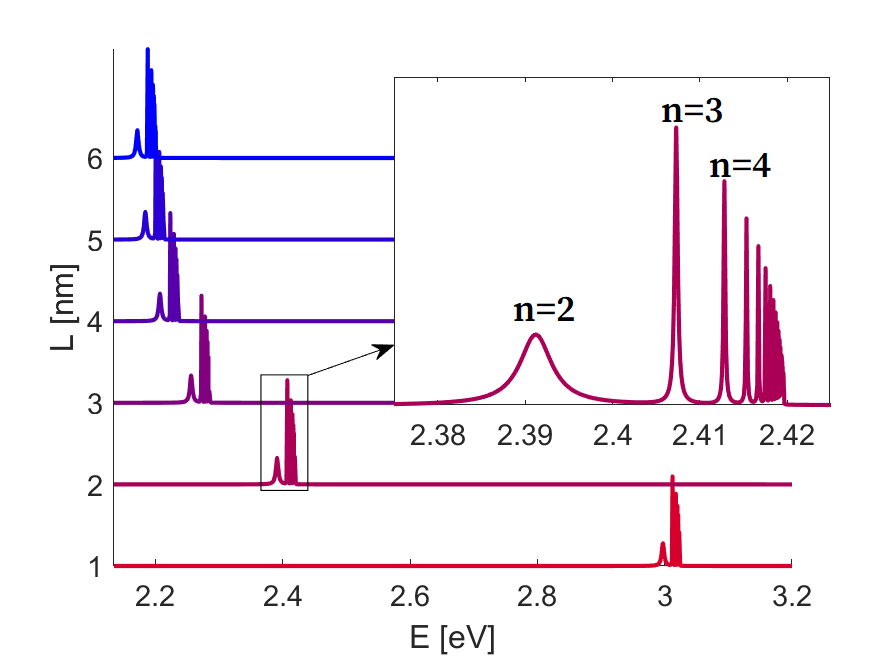}
b)\includegraphics[width=.8\linewidth]{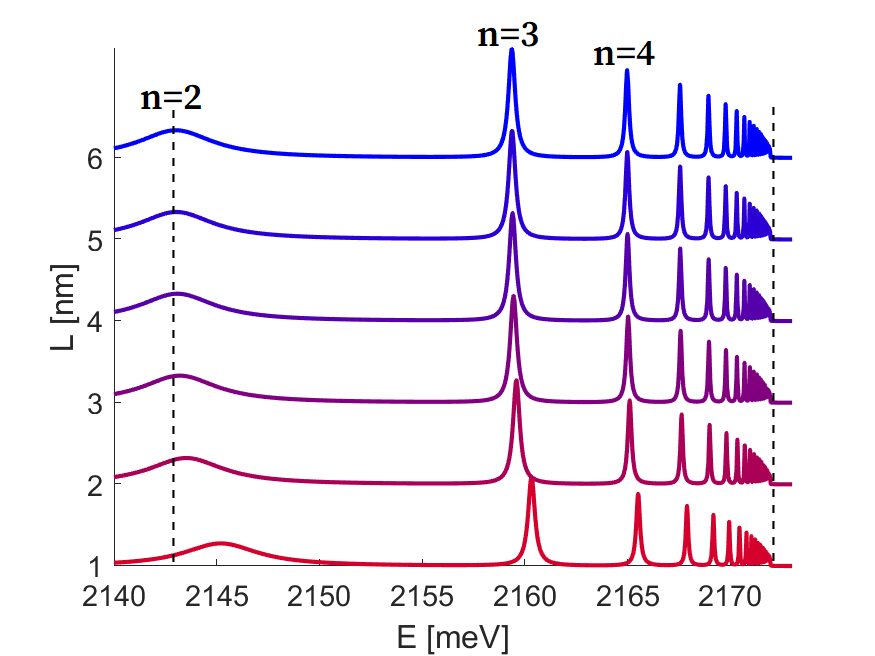}
\caption{Imaginary part of superlattice susceptibility a) with and b) without lowest Cu$_2$O/MgO SL band energy shift included. Inset: zoom on excitonic spectrum for L=2 nm.}\label{rys:2}
\end{figure}

As the SL period L is decreased, the energy shift increases proportionally to $\sim 1/L$. This is shown in Fig. \ref{rys:2} a). For values $L<4$ nm, the shift exceeds the total width of excitonic spectrum which means that the confinement energy exceeds the Rydberg energy. It should be stressed that the energy shift of excitonic spectrum is very considerable, exceeding 1 eV for $L<1$ nm. 
By omitting the energy shifts $E_{e0}$, $E_{h0}$ (Fig. \ref{rys:2} b), one can see the smaller effects; As $L$ decreases, there is a slight decrease of Rydberg energy; one can see this by comparing spectra in Fig. \ref{rys:2} b) with dashed, vertical lines that mark n=2 exciton energy and gap energy for $L=6$ nm, which is close to bulk. For small values of L, the excitonic spectrum becomes visibly narrower, which is a result of smaller effective Rydberg energy $\eta^2_{\ell m}(\alpha)R^*$.

\begin{figure}[ht!]
\includegraphics[width=.8\linewidth]{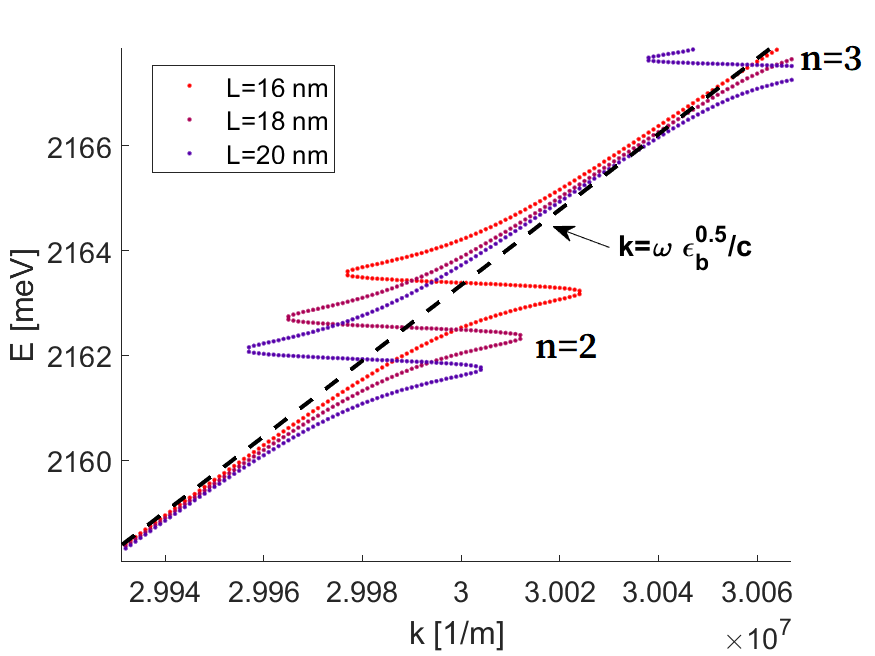}
\caption{Polariton dispersion relation of Cu$_2$O/MgO superlattice, calculated from Eq.(\ref{dispersion}), for three values of SL period $L$, in the energy region of n=2 exciton. Dashed line marks dispersion relation calculated for $\epsilon_b=7.5$, without excitonic effects.}\label{rys:3}
\end{figure}

As a next step, one can calculate the dispersion relation from Eq. (\ref{dispersion}). The results for 3 values of $L$ ($L_W=L_B=L/2$) are shown in Fig. \ref{rys:3}.
For clarity, an energy region near n=2 exciton is chosen; similar shape of the function $E(k)$ is present for every excitonic resonance (n=3 is visible in upper right corner). Overall, the excitons result in a small, localized disturbance of the bulk dispersion relation (dashed line) which is energy shifted depending on the value of L.

\begin{figure*}[ht!]
\includegraphics[width=.7\linewidth]{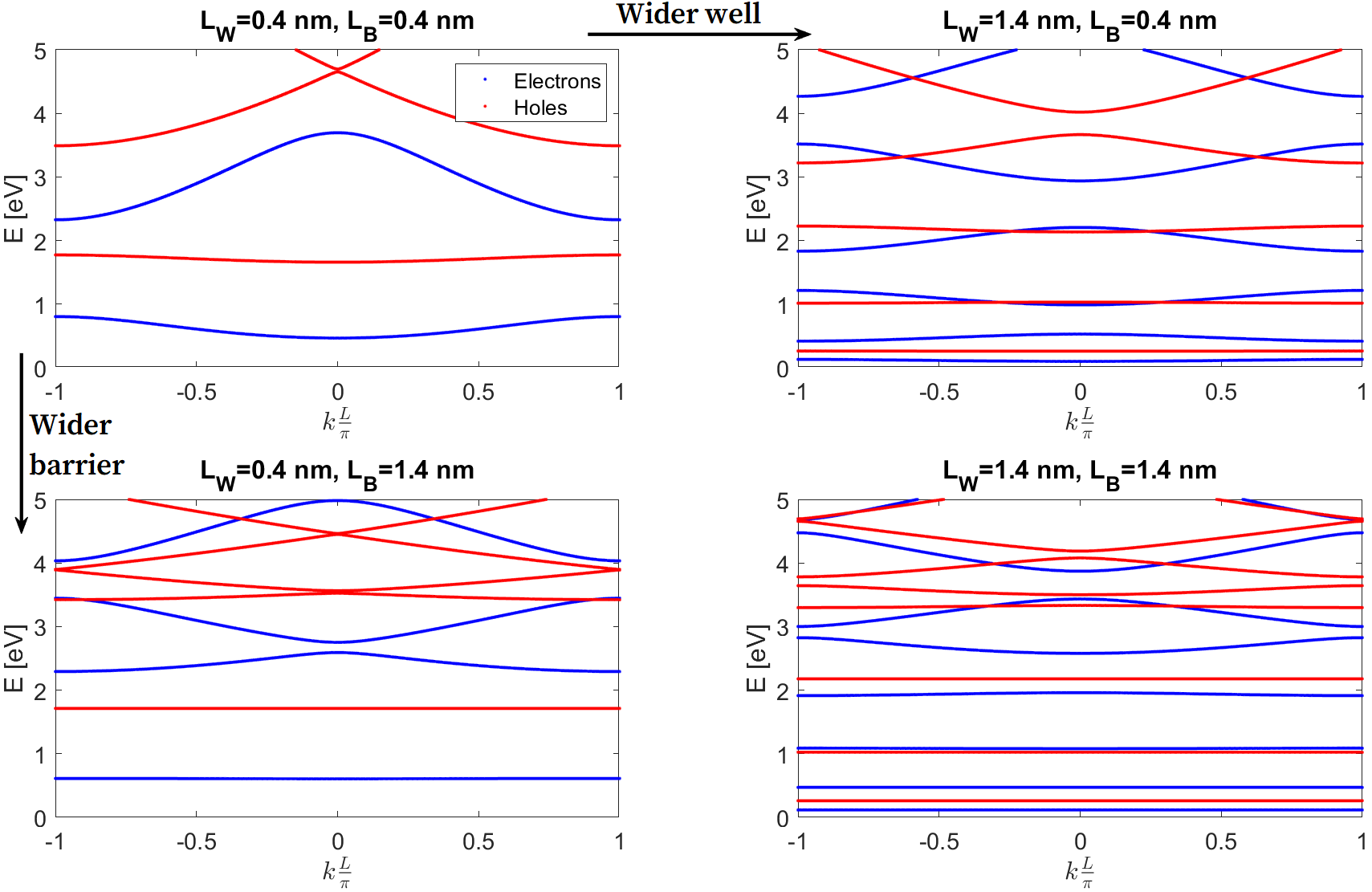}
\caption{SL Dispersion relations of electrons and holes, calculated from Eq. (\ref{KPbound}) for various values of well ($L_W$) and barrier ($L_B$) widths.} \label{rys:4}
\end{figure*}

To calculate the relevant quantities, such as electron and hole effective masses in the $z$ direction and the anisotropy parameter $\alpha$, one has to solve numerically the Eq. (\ref{KPbound}). The results obtained for a few selected values of $L_W$, $L_B$ are presented in Fig. \ref{rys:4}.
For the smallest well and barrier widths, approximately equal to a single atomic layer, only two electron/hole band pairs are visible in the energy range $E<5$ eV. The lowest band is characterized by a positive effective mass, while the masses in the second band are negative and relatively small ($\partial^2 E/\partial k^2 \ll 0$). The increase of well thickness and barrier thickness both result in a higher density of bands, although the effect of increased $L_B$ is less significant. Notably, the dispersion relation of the lowest band becomes extremely flat, especially for holes ($\partial^2 E/\partial k^2 \rightarrow 0$), which results in a very big hole effective mass in the $z$ direction. 

\begin{figure}[ht!]
\includegraphics[width=.7\linewidth]{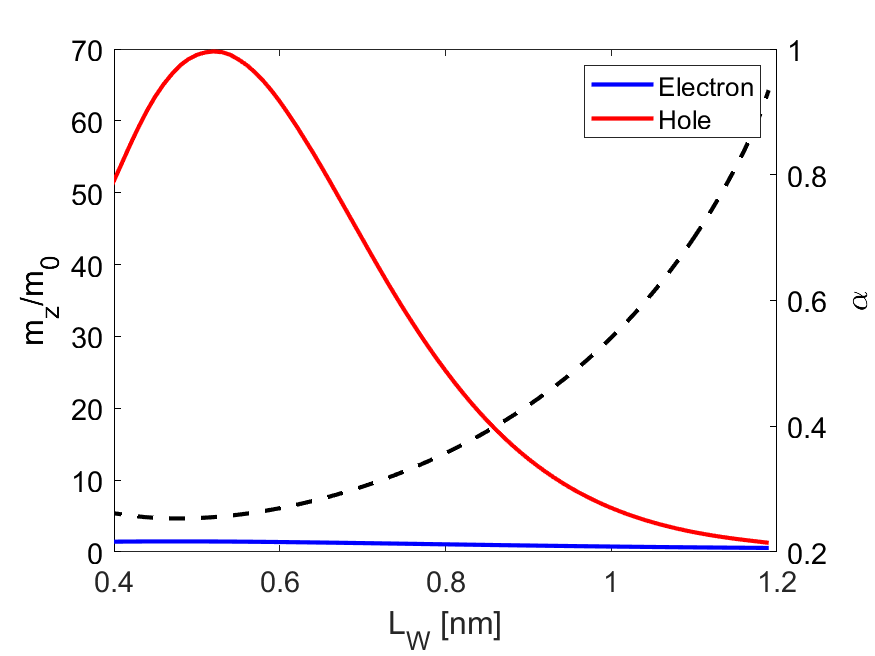}
\caption{Electron and hole effective masses (Eq. (\ref{eq:masy}), left axis) and anisotropy parameter $\alpha$ (calculated from Eqs. (\ref{mureduced}) and (\ref{energies}), right axis) of Cu$_2$O/MgO SL, as a function of $L_W$; $L_W+L_B=1.4$ nm.}\label{rys:5}
\end{figure}

The Fig. \ref{rys:5} depicts the effective electron and hole masses (Eq. (\ref{eq:masy}) as well as anisotropy parameter $\alpha$ (Eqs. (\ref{mureduced}) and (\ref{energies})) as a function of $L_W$, where $L_W+L_B=1.4$ nm.
One can see that there is some optimal value of $L_W=0.5$ nm, $L_B=1.5$ nm, where effective masses are maximized the anisotropy parameter reaches minimum value $\alpha \approx 0.25$ for a slightly smaller well width. As mentioned above, the effective mass of the hole can reach a very high value in the considered system, up to $m_z \sim 70$ $m_0$, while the effective electron mass does not exceed 3 $m_0$. Significantly increased effective mass in $z$ direction means that the system approaches a quasi two-dimensional one, with only two degrees of freedom ($x,y$) for exciton motion. This is reflected in a small value of anisotropy parameter $\alpha$.

\begin{figure}[ht!]
\includegraphics[width=.9\linewidth]{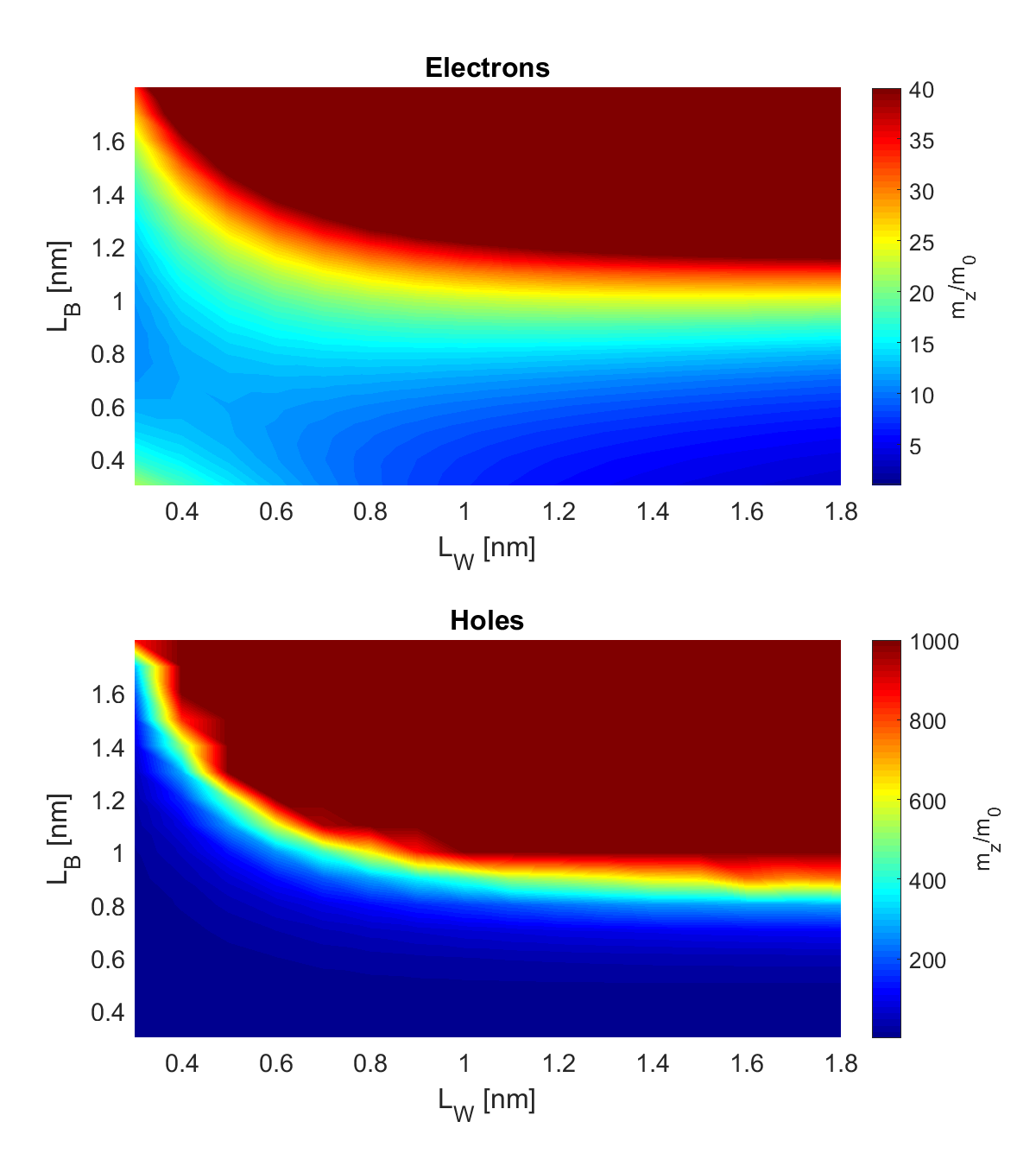}
\caption{Electron and hole effective masses in Cu$_2$O/MgO SL as a function of $L_W$ and $L_B$, calculated from Eq. (\ref{eq:masy}).}\label{rys:6}
\end{figure}

An overview of effective mass values for a range of $L_W$, $L_B$ is shown in Fig. \ref{rys:6}.
In the limit of wide barriers, the effective mass of a hole can reach values of up to $10^3$ $m_0$. In practice, this means that the hole cannot tunnel through the potential barrier and the system becomes two dimensional, allowing only for the motion in $xy$ plane; in such a case, the structure is no longer a superlattice, but a set of separated quantum wells (a multi-well system). The obtained results confirm that the barrier width should not considerably exceed the Bohr radius of the exciton (1.1 nm). It should be stressed that the small barrier width is a necessary condition for the tunelling to occur, which is needed for validity of the presented approach. Another effect visible in Fig. \ref{rys:6} is that in the case of a narrow well ($L_W<1$ nm), the increase of effective mass is slower due to the fact that the well thickness is smaller than exciton diameter, so that its wavefunction enters the barriers, facilitating easier tunneling through them.

\begin{figure}[ht!]
\includegraphics[width=.9\linewidth]{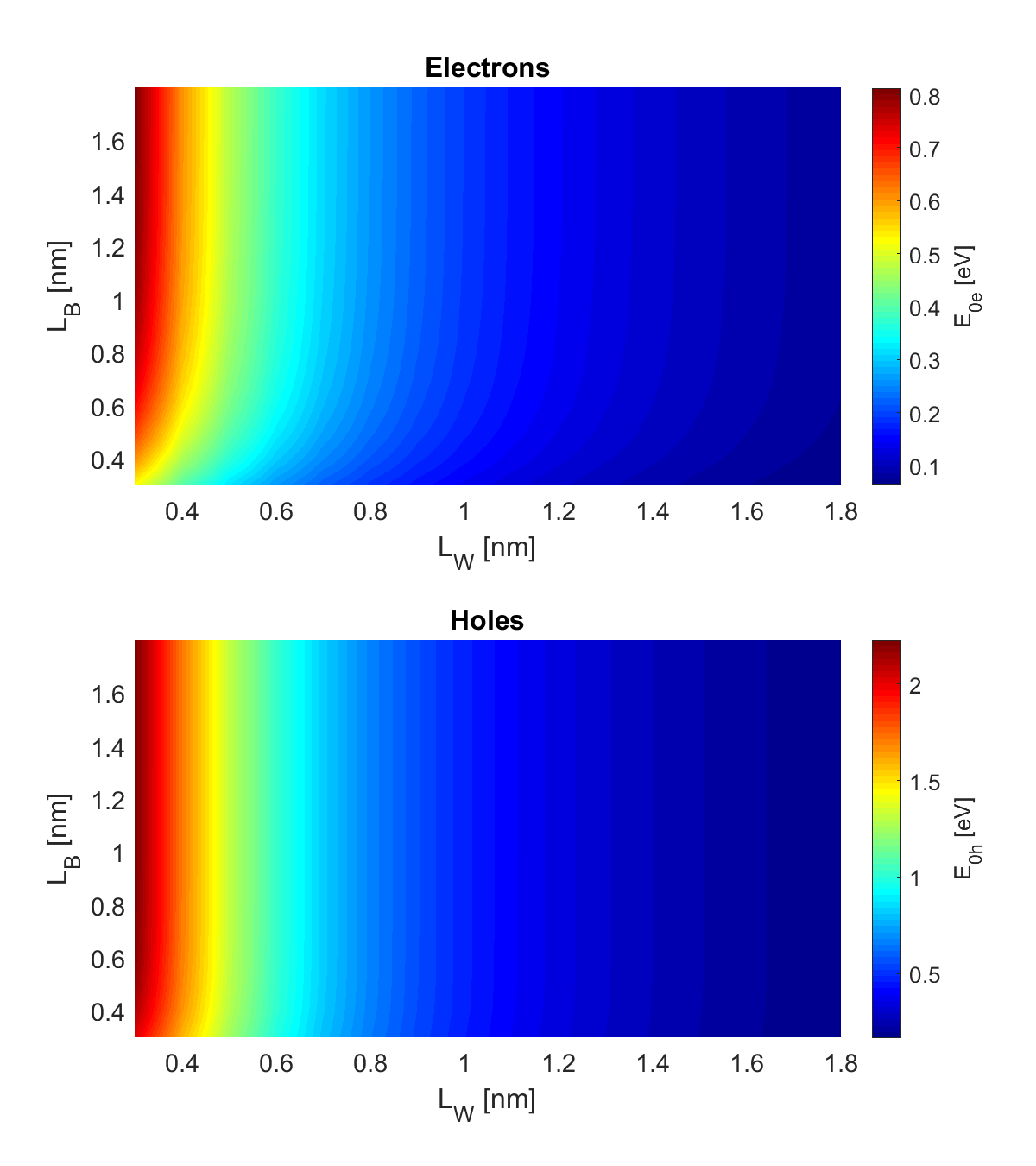}
\caption{The energy of the lowest electron and hole band in Cu$_2$O/MgO SL, as a function of $L_W$ and $L_B$, calculated from  Eq. (\ref{KPbound}).}\label{rys:7}
\end{figure}

Fig. \ref{rys:7} depicts the lowest polariton band energy as a function of $L_W$, $L_B$. 
One can see that the energy dependence on well width is much more pronounced, resulting in $E_{0e} \sim 0.8$ eV and $E_{0h} \sim 2.2$ eV for $L_W = 0.4$ nm. This result is analogous to the case of quantum well, where the energy of the lowest level strongly depends on the well width.

\begin{figure}[ht!]
\includegraphics[width=.9\linewidth]{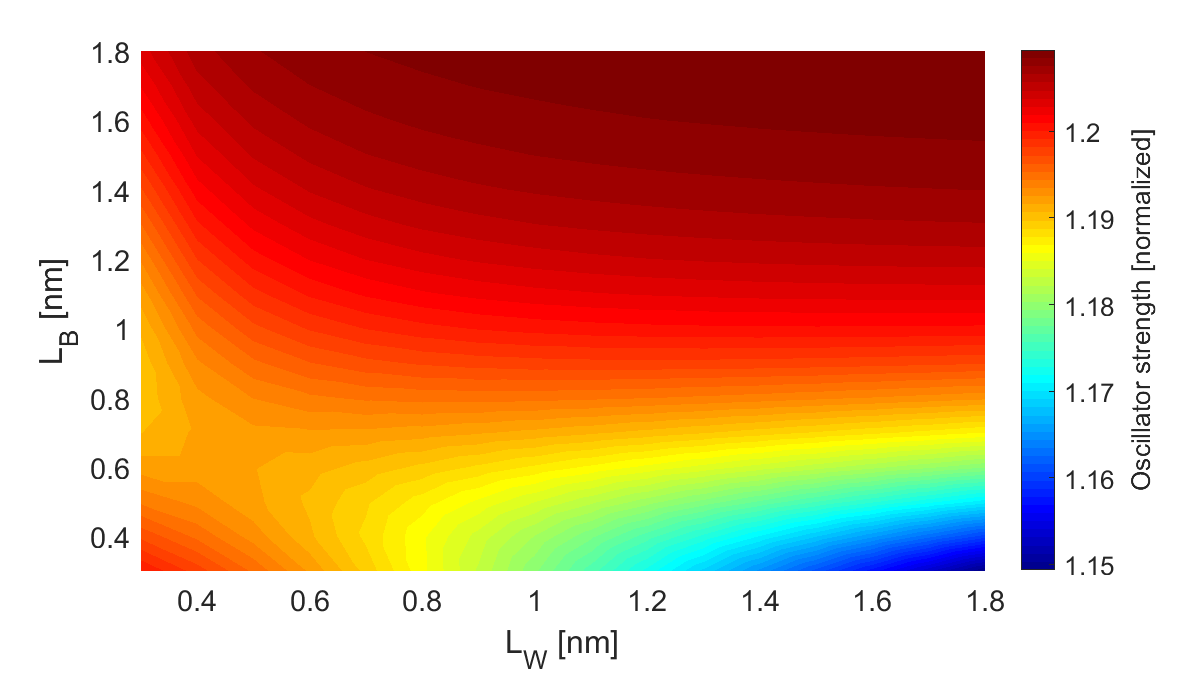}
\caption{Exciton oscillator strength in Cu$_2$O/MgO SL, as a function of $L_W$ and $L_B$, calculated from  Eq. (\ref{eq:fn11}), normalized to the bulk value.}\label{rys:8}
\end{figure}

Finally, in Fig \ref{rys:8} one can see that the effective oscillator strength of the excitons is slightly enhanced in SL, in particular when both barrier and well thickness is large. 
This is expected result - as the effective masses $m_{ez}$m $m_{hz}$ and increase, the reduced mass $\mu_z$ (Eq. (\ref{mureduced})) also increases and anisotropy parameter $\alpha$ (Eq. (\ref{energies})) is decreasing. This, according to Eq. (\ref{etaellm}), affects the oscillator strength. 

\begin{figure}[ht!]
\includegraphics[width=.7\linewidth]{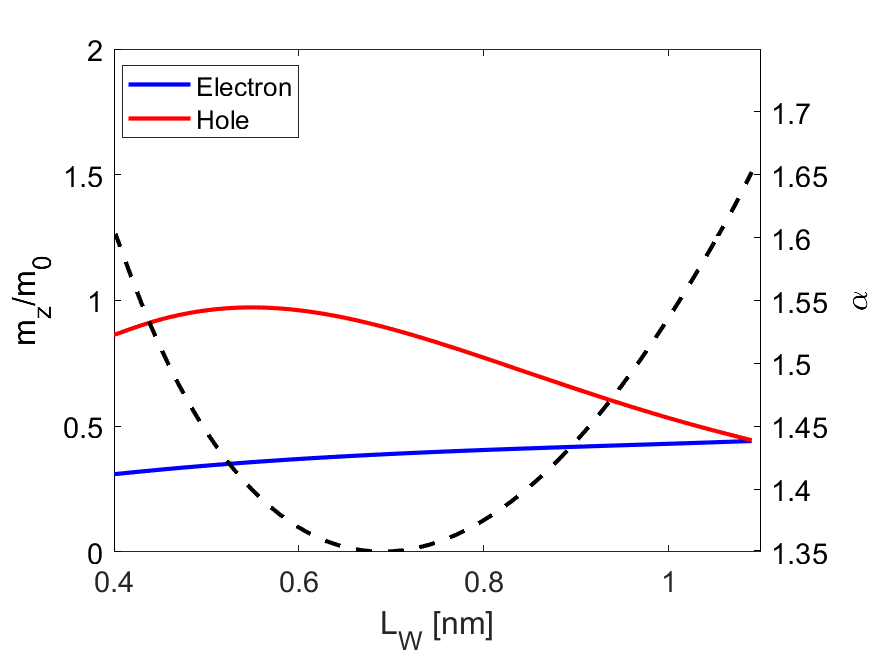}
\caption{Electron and hole effective masses (Eq. (\ref{eq:masy}), left axis) and anisotropy parameter $\alpha$ (calculated from Eqs. (\ref{mureduced}) and (\ref{energies}), right axis) of Cu$_2$O/ZnO SL, as a function of $L_W$; $L_W+L_B=1.4$ nm.}\label{rys:9}
\end{figure}

Superlattice containing Cu$_2$O can use various barrier materials; one of the possibilities is ZnO \cite{Siol2016}. In contrast to MgO, ZnO is a semiconductor with relatively narrow band gap $E_g=3.4$ eV \cite{Norton}, which results in barrier energies of $E_{0e}=0.7$ eV and $E_{0h}=1.88$ eV. The relatively small value of the barrier yields low effective masses in z direction and a large value of $\alpha$. Results of calculations are shown in Fig. \ref{rys:9}. The material parameters used in calculations are given in Table. \ref{parametervalues}.

\begin{table}[ht!]
\caption{\small Parameter values for bulk Cu$_2$O, MgO and ZnO; masses in free electron mass
$m_0$, $R^*$ calculated from $(\mu/\epsilon_b^2)\cdot
13600\,\hbox{meV}$,\, $R^*_{e,h}=(m_{e,h}/\mu)R^*,^{(a)}$
calculated by the assumption that the masses in the $x-y$ plane
remain unaltered, \hbox{lengths in nm},
$a^*_{e,h}=(\mu/m_{e,h})a^*$}
\begin{center}
\begin{tabular}{p{.17\linewidth} p{.17\linewidth} p{.17\linewidth} p{.17\linewidth} p{.25\linewidth}}
\hline
Parameter & Cu$_2$O (Bulk) & MgO (Bulk) & ZnO (Bulk) & References\\
\hline $E_g$ & 2172.08 & 7160 & 3400 & \cite{Kazimierczuk,Yan2012,Norton}\\
$R^*$&87.78& & 60 & \cite{Norton} \\
$\Delta_{LT}$&$0.0125$&   &   & \cite{Stolz}\\
$m_{ez}$ & 0.99 & 0.378 & 0.24 & \cite{Naka,Miller2007,Wang2016,Norton}\\
$m_{h\parl}$ & 0.58 & 1.575 & 0.54 & \cite{Naka,Miller2007,Wang2016,Norton}\\
$m_{hz}$ & 0.58 & 1.575 & 0.54 & \cite{Naka,Miller2007,Wang2016,Norton}\\
$\mu_{\parl}$ & 0.363 & 0.319 & 0.17 & \\
$\mu_z$ & 0.363 & 0.319 & 0.17 & \\
$M_{z}$&1.56&1.953& 0.83 &\\
$\alpha$&1&1&1&\\
$a^*$&1.1& & &\cite{Kazimierczuk}\\
$r_0$&0.22& & &\cite{Zielinska.PRB}\\
$\epsilon_b$&7.5 & &8.656&\cite{Kazimierczuk,Norton}\\
$\Gamma_j$&3.88/$j^3$ & & &\cite{Kazimierczuk,DZ_OL}\\
\hline
\end{tabular} \label{parametervalues}\end{center}
\end{table}

\section{Conclusions}
In conclusion, we  have developed a simple mathematical procedure to calculate in analytical form the susceptibility of superlattice  with Rydberg excitons taking as an example Cu$_2$O/MgO SL, in the  case of normal incidence  of the exciting electromagnetic wave. With the help of Kronig-Penney model for superlattice we have calculated effective masses of  a hole and electron and then used real density matrix approach to obtain resonances for any REs and the polariton dispersion relation. Periodic potentials of the SL structure causes the change of effective masses, which results in the increase of oscillator strengths and significantly shifts positions of the excitonic resonances by over 1 eV. This sensitivity of the energy of excitonic resonances to the SL dimensions may provide an efficient way of measuring mechanical deformation and temperature via thermal expansion of the lattice. The influence of the SL geometry, i.e., the well and barrier thicknesses, on an excitonic spectrum and the dispersion relation have been examined. It turned out that the periodic potential of the considered geometry leads to a significant anisotropy and either small or large value of the anisotropy factor $\alpha$, which can be tuned depending on intended structure application. Finally, we note that Rydberg excitons confined in a superlattice may be a promising platform for quantum computing technologies, being a solid-state analogue of an atomic optical lattice, with significant advantage of compactness and higher operating temperatures.

\section{Acknowledgments}
Support from the National Science Centre, Poland (NCN), project
Miniatura, 2022/06/X/ST3/01162, is greatly acknowledged.

\end{document}